\documentclass{aastex631}

\usepackage{CJK}
\usepackage{color}

\def\cmsq{\hbox{cm$^{-2}$}}
\def\kmps{\hbox{km $\rm{s^{-1}}$}}
\def\nv{N~{\sc v}}
\def\civ{C~{\sc iv}}
\def\siiv{Si~{\sc iv}}

\def\oiii{O~{\sc iii}}
\def\feii{Fe~{\sc ii}}

\def\oi{O~{\sc i}}

\shorttitle{Warm Absorbers in PG 1001+054}
\shortauthors{Fu et al.}
\graphicspath{{./}{figures/}}

\begin{document}
\begin{CJK*}{UTF8}{gbsn}
\title{Fast Outflowing Warm Absorbers in Narrow-Line Seyfert 1 Galaxy PG 1001+054 Revealed by HST/COS Spectra}

\author{Xiao-Dan Fu}
\affiliation{Department of Astronomy, Xiamen University, Xiamen, Fujian 361005, China}

\author[0000-0003-4874-0369]{Junfeng Wang}
\affiliation{Department of Astronomy, Xiamen University, Xiamen, Fujian 361005, China}

\author[0000-0003-0970-535X]{Xiaoyu Xu}
\affiliation{Department of Astronomy, Xiamen University, Xiamen, Fujian 361005, China}

\author{Zhi-Xiang Zhang}
\affiliation{Department of Astronomy, Xiamen University, Xiamen, Fujian 361005, China}

\correspondingauthor{Junfeng Wang}
\email{jfwang@xmu.edu.cn}



\begin{abstract}

Narrow-Line Seyfert 1 (NLS1) Galaxies are an important type of active galactic nucleus (AGN), generally expected to be accreting at high Eddington rate. The properties of their outflows and importance of AGN feedback remain intriguing. We report on the discovery of fast outflowing warm absorbers (WAs) in the NLS1 PG 1001+054, with velocities in the range of $\sim$7000-9000 km s$^{-1}$. They are identified with blueshifted Ly$\alpha$, \nv\ and \siiv\ lines in the high resolution ultraviolet (UV) spectra taken with the Cosmic Origins Spectrograph (COS) onboard the Hubble Space Telescope (HST). We perform photoionization modeling using XSTAR with three WAs. The derived physical properties are typical of WAs in terms of ionization and column density, whereas the outflow velocities are significantly higher. The estimated location of these WAs ranges from 1 to 73 parsecs away from the AGN. Together with previous detection of high ionization absorber in the X-ray for PG 1001+054, we suggest that the fast outflowing UV absorber is probably a part of a multiphase outflow. Such structure is likely produced by the outflow launched from AGN at accretion disk scale, which shocks the ambient ISM producing stratified absorbers.   Assuming contribution from the three WAs at tens of parsecs, the estimated ratio between the kinetic power of the outflow and AGN Eddington luminosity could reach 1.7$\%$, raising the possibility of sufficient influence on the host galaxy when compared to some theoretical models for efficient AGN feedback.


\end{abstract}


\keywords{Ultraviolet sources (1741) --- Quasars (1319) --- Broad-absorption line quasar (183) --- Seyfert galaxies (1447)}


\section{Introduction} \label{sec:intro}

Active Galactic Nucleis (AGNs) are powered by the accretion of gas onto supermassive black hole (SMBH), and AGN driven feedback is generally considered important in the co-evolution between AGN and its host galaxy over the cosmological timescale \citep[e.g.,][]{2012ARA&A..50..455F}. The physical processes of feedback that could have an impact on the interstellar medium and the intergalactic environment remain to be one of the most intensively studied subjects in the past two decades. 

Outflows, either in the form of winds or jets, are the important carrier of the energy output in feedback and found to be prevalent in AGNs \citep[e.g.,][]{2015ARA&A..53..115K,2019ARA&A..57..467B,2020A&ARv..28....2V,2022MNRAS.513.4208S}. They are common and frequently observed from blue-shifted absorption lines both in UV and X-ray bands (\citealt{Laha+21}; \citealt{Bu+etal+2021}; \citealt{2022MNRAS.517.1048B}).It is currently believed that the possible origins of the outflow include accretion disk wind (\citealt{Elvis+etal+2000}), torus wind (\citealt{Krolik+etal+2001}), board-line region (hereafter BLR) clouds (\citealt{Risaliti+etal+2010}) and narrow-line region (hereafter NLR) clouds (\citealt{Kinkhabwala+etal+2002}). The velocity of the outflow is usually several hundred to several thousand kilometers per second (\citealt{McKernan+etal+2007}), hence it can propagate to a location of about 10 kpc (assuming a $\sim$10 million-year lifetime of the AGN) away from the central black hole. In order to investigate the feedback efficiency of the AGN, it is necessary to quantify the physical properties of the outflows (\citealt{Meena+etal+2021}).

Sometimes the inferred velocity of the outflows detected via absorption of highly ionized metal lines can reach tens of thousands of kilometers per second. Such outflows are defined as ultra-fast outflows (UFOs) (\citealt{2002ApJ...579..169C,2003ApJ...593L..65R,Tombesi+etal+2010}), which are likely launched at the scale of innermost accretion disk. The mass outflow rates caused by these outflows can reach dozens of solar mass per year \citep{Tombesi+etal+2013}, implying large kinetic energies as expected in theoretical predictions of AGN feedback models \citep[e.g.,][]{Hopkins+etal+2010,King+10,2015ARA&A..53..115K}.  Such energetic outflows may suppress the star formation in host galaxy, prevent the gas around the galaxy from cooling (\citealt{Krongold+etal+2007}) and increase the abundance of intergalactic filaments (\citealt{Hopkins+etal+2010}). 

Warm absorbers (WAs), characterized by narrow absorption lines and photo-absorption edges, also reveal the presence of the ionized phase outflow. They are detected in about 50$\%$ of Type I Seyfert galaxies either in the ultraviolet (UV) or in the X-ray band \citep{Reynolds+etal+1997,Crenshaw+etal+1999,Blustin+etal+2005}. The WAs detected in the soft X-ray typically have higher ionization states compared to their UV counterparts \citep{2013MNRAS.430.2650L,Laha+etal+2014,Fu+17}. Compared to UFOs, the WA outflows typically show much lower velocities in the range of a few 100 to 1000 km s$^{-1}$,  and a kinetic luminosity $\ll 1\%$ of the bolometric luminosity of the AGNs \citep{Blustin+etal+2005,Tombesi+etal+2013}.

UV spectra typically have high spectral resolution allowing accurate dynamic measurement, and cover line transitions of ions in low ionization states. In contrast, X-ray spectra cover more transitions from ions at various ionization states, but are subject to lower resolution and signal to noise (S/N) ratio (e.g. \citealt{Kaspi+etal+2002}). It is challenging to establish the one-to-one correspondence for the outflows between UV and X-ray band. The absorption column densities of outflows in these two bands can differ by hundreds of times (e.g. \citealt{Ulrich+etal+1988}). As a result, outflows in these two bands may only partially overlap (\citealt{Kriss+etal+2004}). Taking into account the rapid variations in luminosity, the approach adopted in recent years is to conduct simultaneous observations of certain sources \citep[e.g. Mrk 509, Mrk 279, NGC 4051, 1H0419-577, NGC 3227;][]{Costantini+etal+2007,Costantini+2010,Kriss+etal+2011,2013A&A...556A..94D}.  A broad understanding of these WAs in the two bands has yet to emerge \citep{Crenshaw+Kraemer+George+2003}.

The possible connection among the various type of ionized outflows was investigated in detail in \citet{Tombesi+etal+2013}, where the comparison between the UFOs and the WAs suggests both types of ionized outflows belong to a single stratified outflow. Recent studies using high-resolution grating have detected UFOs in the soft X-rays \citep[e.g., Ark 564, PDS 456, PG 1211+143, IRAS 17020+4544, Mrk 1044;][]{2013ApJ...772...66G,2015Sci...347..860N,2016MNRAS.459.4389P,2020ApJ...895...37R,2015ApJ...813L..39L,Krongold+etal+2021}, typically with lower column densities and ionization parameters compared to that of the UFOs identified in the 6--9 keV range. Interestingly, the high-velocity UV counterparts to some of these UFOs have also been reported \citep{2012ApJ...751...84K,Kriss+18,2018MNRAS.476..943H,M+22}, showing narrow and blueshifted absorption lines (e.g., HI Ly$\alpha$, \nv, \civ\, etc.). Altogether these strengthen the conjecture that they could be part of the same stratified, multi-ionization outflow. Previous work further discovered a correlation between the wind outflow velocity and the hard X-ray luminosity of the AGN, suggesting that the UFOs could be consistent with a predominately radiatively driven wind \citep{Matzeu+17,2021MNRAS.503.1442M}, arising in systems accreting at or close to the Eddington rate.  Nevertheless, such physical relationship and the launching mechanism still remain unclear \citep[see review by][]{2020IAUS..342...90L,Laha+21}.  

Over the past decade, HST/COS has performed highly sensitive high resolution observations towards UV bright quasars, yielding abundant information of the intervening absorption systems, such as the gaseous halos and the circumgalactic medium (CGM) of galaxies along the sight-line to the quasar \citep[e.g.,][]{2013ApJS..204...17W,2016A&A...590A..68R,2017ARA&A..55..389T,2021ApJ...923..189M}. Often spectral features associated with the background quasars were not fully analyzed. This under-explored collection of archival spectra is a treasure trove for carrying out WAs, UFOs, and AGN outflow studies in general, which motivated our work. 

A peculiar subgroup of AGNs characterized by high Eddington ratios \citep{1992ApJS...80..109B,2000NewAR..44..431P} are Narrow-Line Seyfert 1 (NLS1) galaxies. Their other typical features include: (1) the full width at half maximum (FWHM) of hydrogen Balmer lines less than 2000 ~\kmps (\citealt{Turner+etal+1999}); (2) the line ratio of [\oiii]~$\lambda 5007~\textrm{\AA}$ to H$\beta$ less than 3 (\citealt{Leighly+etal+1999}); (3) high line ratio of \feii~to H$\beta$ (\citealt{Mathur+etal+2000}); (4) generally steeper soft X-ray continuum slopes compared to other Seyfert 1 galaxies and rapid soft X-ray variability. (\citealt{Boller+etal+1996}); (5) strong infrared emission indicating active star formation (\citealt{Moran+etal+1996}). Indeed, several UFOs with high velocity X-ray and UV absorbers are reported in NLS1s (e.g., NGC 4051, \citealt{2012MNRAS.423..165P}; IRAS 17020+4544, \citealt{M+22}).

The NLS1s with ionized outflows were selected based on the catalog of \citet{Rakshit+etal+2017}, which provides a list of 11101 NLS1 galaxies identified from the Sloan Digital Sky Survey Data Release 12. This catalog is about 5 times larger than the number of previously known NLS1 galaxies. We retrieved the {\sl HST}/Cosmic Origins Spectrograph (COS) observation of these sources and obtained a total of 72 NLS1s with UV observations. We identified one object with unpublished COS spectra that show intriguing high velocity absorption features, PG 1001+054, for a pilot study of the WA from the UV perspective.

PG 1001+054 (z=0.16012, from NED\footnote{http://ned.ipac.caltech.edu}) is classified as NLS1 galaxy, with a FWHM of H$\beta$ $\sim$1740~\kmps (\citealt{Wills+etal+2000}). Meanwhile, the broad \civ~ and Ly$\alpha$ absorption lines in UV spectrum (a also make it qualified as a broad absorption line (BAL) quasar \citealt{Brandt+etal+2000}; \citealt{Wills+etal+2000}; \citealt{Wang+etal+2000}), a subclass with extremely weak X-ray emission. \citeauthor{Wang+etal+2000} (2000) found that the observed UV line optical depth is much lower than expected from the X-ray absorbing column density in this source, based on a comparison between the {\em ROSAT} soft X-ray detection and the UV BALs.  Despite the low S/N ratio, the analysis of its {\em XMM}-Newton EPIC spectra \citep{Schartel+etal+2005} found evidence for absorption through ionized material, modeled with a column density $N_{\mathrm{H}}$ of $19.2\times10^{22}$~\cmsq and an ionization parameter $\xi$ of 542 $\rm~{erg~s^{-1}~cm}$, respectively. Recent {\em NuSTAR} observations \citep{2022ApJ...936...95W} investigated the nature of its X-ray weakness, and suggested that X-ray obscuration by clumpy dust-free wind is sufficient to explain the variation of multi-epoch X-ray data and the X-ray weakness.

In this work, we present the first detailed spectroscopic analysis and photoionization modeling of the high-resolution COS spectra of PG 1001+054. We identify high velocity WAs ($v\sim 6700-8900$ \kmps), which demonstrates that high-resolution UV spectroscopy with HST can play a crucial role in feedback studies providing physical information such as kinematics and ionization structure about the fast outflowing gas. Section \ref{sec:reduct} describes the observation and data reduction. In Section \ref{sec:analy} we analyze the emission and absorption components in the spectra and prepare the suitable models for them, using the same procedure in \citet{Zhang+etal+2015} and \citet{Fu+17}. We present the fitting results in Section \ref{sec:fitting}. In Section \ref{sec:discuss} we discuss and interpret the results of our modeling, and finally Section \ref{sec:summary} gives a brief summary of our conclusions.

\section{Observations and data reduction} \label{sec:reduct}

PG 1001+054 was observed by {\em HST}/COS in June 2014 (PI: J. Bregman) with an exposure time of 7 ks. It was observed as a target in a large sample to study missing baryons in nearby dwarf galaxies, and has not been studied in detail individually. The log of the observations is listed in Table \ref{tab:tab1}. The COS observations consist of eight segments utilizing two gratings, G130M and G160M, which are centered at 1300\AA~and 1600\AA, respectively. This yields spectra fully covering a wavelength range of 1135\AA~to 1795 \AA~(\citealt{Osterman+etal+2011}), with a resulting medium spectral resolution $R\equiv\lambda/\Delta\lambda$ from 16000 to 21000. 

\begin{table}
\begin{center}
\caption{The log of the HST/COS observations of PG 1001+054
\label{tab:tab1}}
\footnotesize
\begin{tabular}{cccc}
\hline\noalign{\smallskip}
 Observation Log  & Start Time       &Stop Time   &Gratings\\
 \hline\noalign{\smallskip}
     & 2014/6/18 22:06:35  & 2014/6/18 22:24:19 &G160M\\
     & 2014/6/18 23:24:15  & 2014/6/18 23:42:09 &G160M\\
     & 2014/6/18 0:59:50   & 2014/6/18 1:14:00 &G160M\\
 COS & 2014/6/18 1:18:10   & 2014/6/18 1:32:20 &G160M\\
     & 2014/6/18 2:35:27   & 2014/6/18 2:46:42 &G130M\\
     & 2014/6/18 2:50:05   & 2014/6/18 3:04:05 &G130M\\
     & 2014/6/18 4:11:00   & 2014/6/18 4:24:45 &G130M\\
     & 2014/6/18 4:28:27   & 2014/6/18 4:42:12 &G130M\\
\noalign{\smallskip}\hline
\end{tabular}
\end{center}
\end{table}

\begin{figure}
\centering
\includegraphics[angle=0,width=7in]{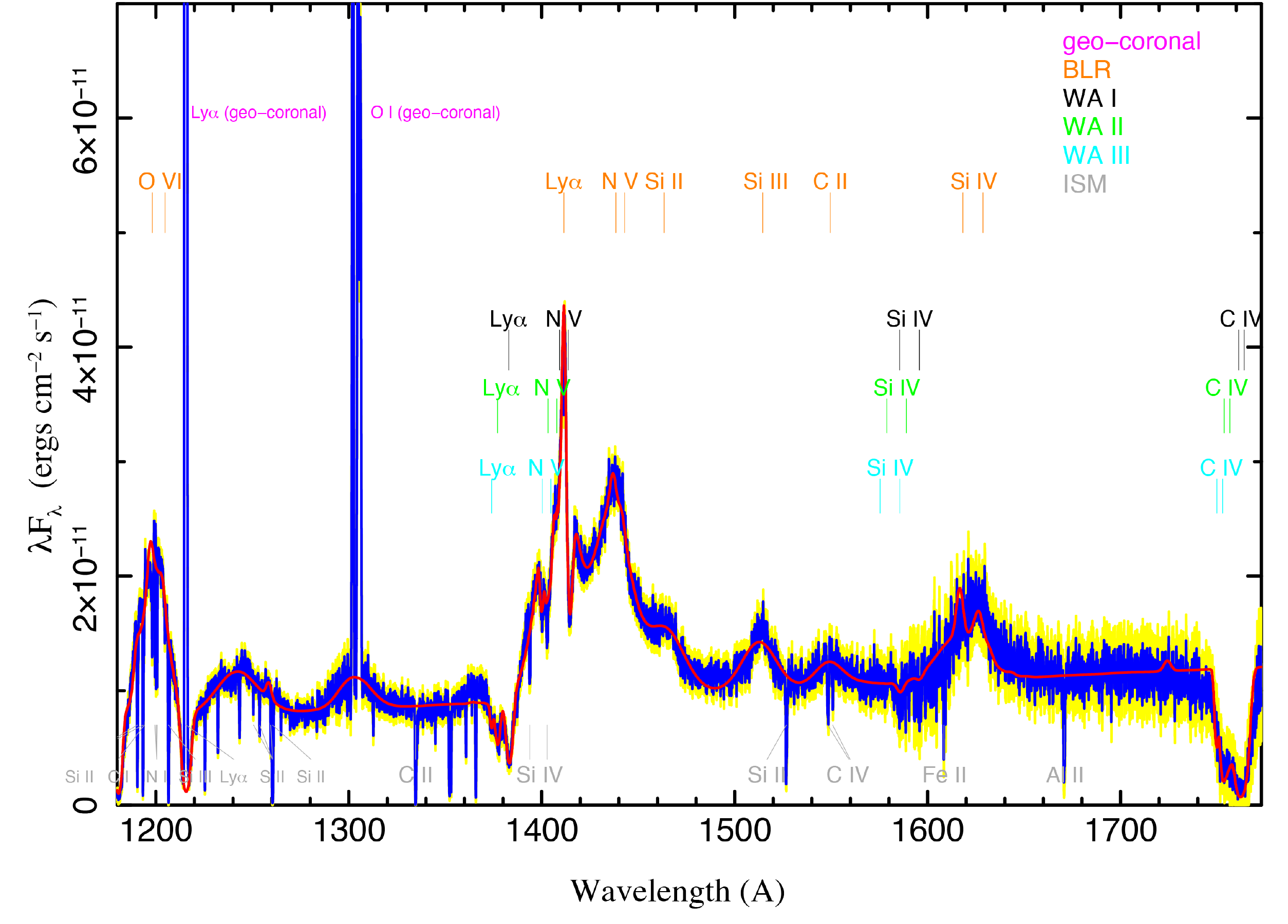}
\caption{The observed COS spectrum of PG 1001+054. The data is shown in blue and the error bar is in yellow. Lines from emission components are labeled with orange notes, and three absorbers of PG 1001+054 are labeled with black, green and light blue notes. The local ISM lines and geo-coronal lines are noted with grey and purple marks, respectively. The best-fit model of continuum, emission and absorption lines is shown in red.
\label{fig:fig1}}
\end{figure}

The archival HST/COS data of PG 1001+054 were retrieved from the Mikulski Archive for Space Telescopes
(MAST\footnote{http://archive.stsci.edu}), fully calibrated and processed with the latest COS calibration pipeline CAL$\mathit{COS}$ (version 3.3.11). The wavelengths of expected local ISM lines and geo-coronal lines are used to verify that the calibrated wavelength is accurate.  We use IDL routines described in \cite{Danforth+etal+2010} to process flat-fielding, alignment, and co-addition. Eight observations are merged with exposure weighting, and the final spectrum’s S/N ratio covers a range of 15 to 25 per resolution element (0.07 \AA, or 17 \kmps). Using the IDL toolkit {\tt Package for Interactive Analysis of Line Emission} \citep[PINTofALE\footnote{ http://hea-www.harvard.edu/PINTofALE/};][]{2000BASI...28..475K}, the COS flux spectrum is finally converted into the format commonly used in X-ray studies. Pulse Height Amplitude (PHA) files and the corresponding response (RSP) files that convolved with G130M and G160M line spread functions (LSF) are generated following the above process. Figure \ref{fig:fig1} presents the final COS spectrum of PG 1001+054.

\section{Spectral Analysis and Modeling}\label{sec:analy}

The processed spectrum is analyzed and fitted by the Interactive Spectral Interpretation System (ISIS\footnote{http://space.mit.edu/cxc/isis/}; version 1.6.2, \citealt{Houck02}), which is a programmable, interactive tool to explore the physics of COS spectrum. 

Taking account of the cosmological redshift of PG 1001+054, we first identify the typical UV emission and absorption lines in the spectrum based on AtomDB\footnote{http://www.atomdb.org} (version 3.0.9) (\citealt{Foster+etal+2012}). Narrow absorption lines caused by the diffuse interstellar medium (ISM) lines have also been identified and masked to exclude interference when fitting the spectrum. Next we utilize the photoionization code XSTAR (Version 2.54a, \citealt{Kallman01}) to model the emitting and absorbing plasma photoionized by the radiation from the central accretion disk. To generate XSTAR model grids, several parameters need to be set, for example, the luminosity of AGN, the hydrogen nucleus density, the metal abundances (usually set to solar values), the spectral energy distribution (SED) file, the temperature, and the turbulent velocity. We set the column density $N_{\mathrm{H}}$ and the ionization parameter $\xi =L_{ion}/(nr^{2})$ as intrinsically free parameters, where $L_{ion}$ is the luminosity in the 1-1000 Ryd energy range, $n$ is the hydrogen nucleus density, and $r$ is the distance to the ionizing source. The parameters for XSTAR table models are listed in Table \ref{tab:tab3}. We use the $b$ parameter in XSTAR to produce the corresponding FWHMs in the observed line profiles, where $b = \rm{FWHM}/2\sqrt{(ln 2)} \approx \rm{FWHM}/1.665$.

\begin{table*}
\begin{center}
\caption{The parameters for XSTAR table models.
\label{tab:tab3}}
\small
\begin{tabular}{lccccccc} \hline\hline
              & $T$          & $n$            & $b\footnote{The Doppler broadening parameter $b$ is approximately equal to FWHM/1.665.}$            & Log $\xi$    &   $N_{\rm{H}}$             & Abundance             & $z$ \\
              & $(\rm{K})$    & $(\mathrm{cm^{-3}})$& $(\kmps)$      &$(\mathrm{erg~s~cm^{-1}})$&$\rm{(\times10^{19}~cm^{-2})}  $      &                      &     \\
\hline\noalign{\smallskip}
\multicolumn{3}{l}{AGN emission} \\
\hline\noalign{\smallskip}
$\rm{BLR~I}$        & 15000          &$10^{10}$       & 2421.22   & $0\sim4$                &$10^{4}$              	        &Solar                 & Free \\
$\rm{BLR~II}$        & 15000          &$10^{10}$       & 642.13   & $0\sim4$                 &$10^{4}$              	        &Solar                 & Free \\
\hline\noalign{\smallskip}
\multicolumn{3}{l}{AGN absorption} \\
\hline\noalign{\smallskip}
$\rm{WA~I}$    &$10^{5}$         &$10^{4}$      & 628.64     & $0\sim3$                &$0.5\sim500$                	                  &Solar   & Free\\
$\rm{WA~II}$   &$10^{5}$         &$10^{4}$        &510.59    &$0\sim3$                 &$0.5\sim500$                 	                  &Solar  & Free\\
$\rm{WA~III}$  &$10^{5}$         &$10^{4}$      & 287.69     &$0\sim3$                &$0.5\sim500$                                &Solar  & Free\\
\hline\noalign{\smallskip}
\end{tabular}
\end{center}
\end{table*}

Considering the UV radiation could be substantially reduced by dust extinction, we obtain extinction corrected radiation using the extinction curve formula proposed by \cite{Gordon+2009}, which covers the wavelength range from 910 $\rm{\AA}$ to 3.3 $\mu$m. Parameters ${R_{\mathrm{v}} =A_V/E(B-V)=3.1}$ (\citealt{Cardelli89}) and ${A_V=0.042}$ (available from NED) are adopted in the case of PG 1001+054. To construct the SED file of PG 1001+054, we collect the available data from NED supplemented by the data from our dust extinction corrected UV spectrum. PG 1001+054 is also known as a bright low redshift quasar, with a bolometric luminosity $L_{bol}=1.35 \times 10^{45}$ erg s$^{-1}$ \citep{2017MNRAS.468.1433P}. The ionizing luminosity of PG 1001+054 is obtained from the SED file as $L_{ion} = 2.14 \times 10^{44}$ erg s$^{-1}$. 

The intrinsic UV radiation of PG 1001+054 generally comes from the accretion disk, BLR, and NLR. A power-law model is used to fit the multiple blackbody emission in the accretion disk empirically. As for the emission from BLR and NLR clouds, we need to generate XSTAR models to fit these components. We take the plausible assumption that the clouds in BLR and NLR have approximately virial speed, i.e. $v\propto \sqrt{{\rm G}M/R}$ (M is the black hole mass and R represents the distance to the center) following the method in our previous work \citep{Zhang+etal+2015,Fu+17}. In this case, the width of emission line in the spectrum is closely linked to the radial velocity dispersion of different groups of photoionized clouds associated with BLR and NLR. Each BLR or NLR component is assumed to have the similar physical conditions such as density, radial distance and temperature, thus can be described by one photoionization component. We caution that recent work on spatially resolved NLR outflows in local AGNs clearly reveals variations in ionization and density \citep{2021ApJ...910..139R}, and we do not further interpret on the model fits of emission line.

These emission lines are first adequately modeled to facilitate identification of absorption systems. Following \citet{Zhang+etal+2015} and \citet{Fu+17}, we model the emission lines using XSTAR. Some previous work adopted other approaches for modeling the emission line, such as multiple gaussians \citep[e.g.,][]{2017ApJS..229...22H,2021ApJ...916...31M}, piecewise function or spline fit \citep[e.g.,][]{2022ApJ...926...60V}. To generate those photoionization components in XSTAR, we fit the strong emission lines with gaussian models to measure the FWHM of each line, and derive the Doppler broadening parameter or turbulent velocity. In the COS spectrum of PG 1001+054, the best candidate lines for profile decomposition are Ly$\alpha$, \nv~doublet, and \siiv~doublet, given their high S/N ratio relative to other lines. The five lines are fitted jointly according to their rest frame wavelengths. The five lines share the same radial velocity and FWHM, and the flux ratio of these doublets is fixed at 2:1 for an optically thin case. Two groups of Gaussians with FWHM values of $\sim$ 4031, 1069 \kmps~ are needed. These quite large FWHM values indicate that the two emission components are likely associated with the BLRs, although $FWHM \sim 1000$ \kmps~ could be possible for a highly turbulent NLR. We tentatively associate the two components as BLR clouds (hereafter BLR1 and BLR2). Two XSTAR table models are generated for them. The density, column density and temperature of the BLR model are set to $10^{10}$ cm$^{-3}$, $10^{23}$ \cmsq and 15000 K, respectively, following the value in \citet{Zhang+etal+2015}. The metallicities are set to the solar values, and we leave the ionization parameter and the redshift as free parameters. The measured redshift by XSTAR will allow us to identify offset from the systemic velocity due to outflow or inflow. After the initial fitting, residual emission represented by one Gaussian component is added to the model. The reduced chi-square of the global best fit is 1.64.  

\section{The Properties of Warm Absorbers} \label{sec:fitting}

\begin{figure}
\centering
\includegraphics[angle=0,width=7in]{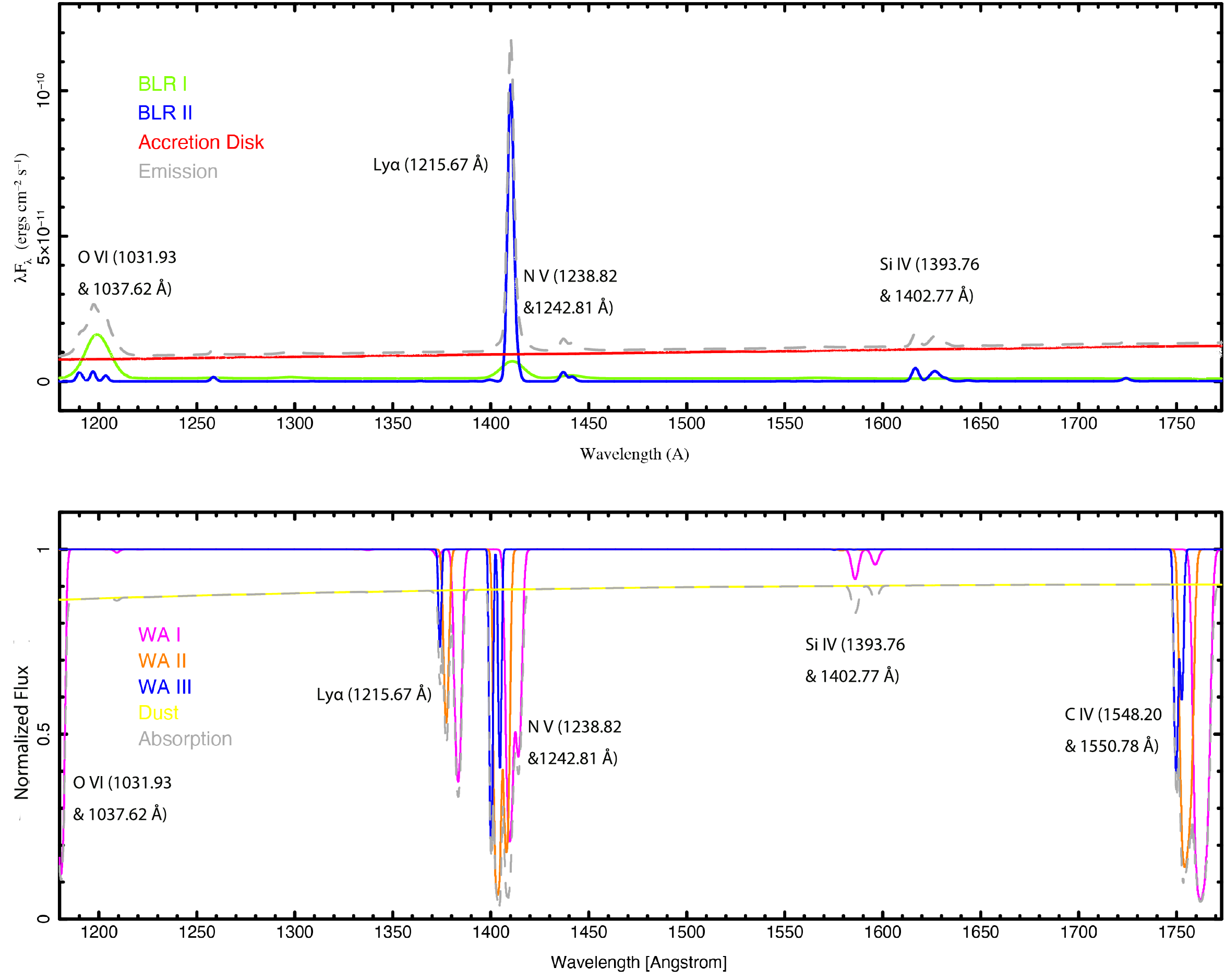}
\caption{Illustration of the separate model components fitted in the PG 1001+054 spectra. Each component is marked with a different color, as labeled. The absorbers here are absorbers without the effect of the covering factor. The gray dashed lines are the combination of emission and absorption components created by the models, respectively.
\label{fig:fig4}}
\end{figure}

We estimate the outflow velocities of WAs through the Ly$\alpha$, \nv~and \siiv~doublet absorption troughs in the COS spectrum. By fitting these absorption lines with Gaussians, we identify three kinematic components (hereafter WA~I, WA~II and WA~III) with blueshifted velocities of $-$8931, $-$6761 and $-$8201 \kmps. The FWHM of these three WAs are about 1047, 850, and 479 \kmps, respectively. Since thermal broadening and turbulent broadening can make a significant impact on the absorption lines. we can calculate the $v_{\rm turb}$ of each WAs according to $\rm{FWHM}=2\sqrt{\rm ln2}\sqrt{v_{\rm th}^2+2v_{\rm turb}^2}$ (\citeauthor{Zhang+etal+2015} 2015). $v_{\rm th}$ is the thermal velocity, which is about $v_{\rm th}=13\sqrt{T/10000A}~\kmps$ where $T$ is the temperature and $A$ represents the atomic number. Typically the temperatures of low ionization WAs are about $10^{5}$ K, and the mean value of carbon and nitrogen atomic numbers 
is adopted as the $A$ value. All of the outflow velocities and turbulent velocity $v_{\rm turb}$ values are listed in Table \ref{tab:tab2}. The separate model components we fit in PG 1001+054 spectra are shown in Figure \ref{fig:fig4}.

\begin{table*}
\begin{center}
\caption{Intrinsic lines fitted with Gaussian in COS spectrum.
\label{tab:tab2}}
\small
\begin{tabular}{ccc|cc|c} \hline\hline
$ Ion (\lambda_{\textsl{rest}})$   & $f\footnote{f is oscillator strength.}$ & $flux\,(\times10^{-4})\footnote{The strong and weak line flux ratio of the doublets are in an optically thin case of 2:1.}$   &   $\lambda_{\textsl{obs}} $      & $Velocity \footnote{Five lines in one component have same radial velocity and FWHM for that they are fitted jointly according to their rest wavelength relations.}$   & $\rm{FWHM}$    \\
$\textrm{\AA}$    &           &$\mathrm{photons^{-1}~cm^{-2}}$  &$\textrm{\AA}$ & $\kmps $ & $\kmps$  \\
\hline\noalign{\smallskip}
\multicolumn{3}{c}{ Emission lines from BLR} \\
\hline\noalign{\smallskip}
Ly$\alpha$ (1215.67)  &0.41617   &84.67  &$1414.03$  &\nodata  &\nodata	     \\
                      &	         &177.33 &$1410.25$  &\nodata  &\nodata	     \\
        
\nv~(1238.82)         & 0.15553  &45.05    &$1441.96^{+0.47}_{-0.25}$   &$914.50^{+112.12}_{-58.58}$   &$4031.33^{+86.81}_{-115.59}$ \\
                      &	         &10.05    &$1437.11\pm0.02$      
&$-17.91\pm5.21$        &$1069.14^{+21.42}_{-9.31}$	             \\
\nv~(1242.81)        & 0.07781   &22.53    &1445.59   &\nodata   &\nodata   \\
                     &	         &5.03  &1441.73      &\nodata    &\nodata    \\
                  
\siiv~(1393.76)     &0.52771    &62.46  &1621.17    &\nodata    &\nodata    \\
                     &           &2.59    &1616.84   &\nodata   &\nodata    \\
                 
\siiv~(1402.77)       &0.26285  &31.23   &$1631.66$ &\nodata   &\nodata   \\
                      &           &1.30   &$1627.30$   &\nodata   &\nodata \\

\hline\noalign{\smallskip}
\multicolumn{3}{c}{ Absorption lines from Warm Absorbers} \\
\hline\noalign{\smallskip}
Ly$\alpha$ (1215.67)  &0.41617   &$14.97$    &1382.89          &\nodata                    &\nodata	                              \\
                      &	         &$726.00$     &1377.06         &\nodata                    &\nodata	                             \\
                      &	         &$2.04$     &1374.11            &\nodata                    &\nodata	                             \\
\nv~(1238.82)   &0.15553         &88.19   &1409.24           &\nodata                    &\nodata	                             \\
                       &	     &21.49   &1403.29          &\nodata                    &\nodata	                             \\
                       &         &7.58    &$1400.28^{+0.03}_{-0.06}$ &$-8931.12^{+7.26}_{-14.52}$  &$479.01^{+41.46}_{-34.44}$  \\
\nv~(1242.81)   &0.07781        &$44.10$ &$1413.77^{+0.04}_{-0.02}$ &$-6761.60^{+9.65}_{-4.82} $  &$1046.68^{+17.99}_{-17.17}$ \\
                       &         &$10.74$ &$1407.80^{+0.03}_{-0.08}$ &$-8201.69^{+7.24}_{-19.30} $ &$850.14^{+34.95}_{-32.71}$  \\
                       &	     &$3.79$  &1404.78    &\nodata                    &\nodata	                             \\
\siiv~(1393.76)   &0.52771      &4.26    &1585.49   &\nodata                    &\nodata	                             \\
                       &	     &0.10     &1578.79  &\nodata                    &\nodata	                           \\
                       &	     &0.10     &1575.40  &\nodata                    &\nodata	                             \\
\siiv~(1402.77)   &0.26285      &2.13    &1595.74   &\nodata                    &\nodata	                          \\
                      &	         &0.05   &1589.00    &\nodata                    &\nodata	                           \\
                      &	         &0.05   &1585.59    &\nodata                    &\nodata	                         \\
\hline\noalign{\smallskip}
\end{tabular}
\end{center}

\end{table*}

We adopt the typical value of the density and temperature in these WAs as $10^{4}$ cm$^{-3}$ and $10^{5}$ K \citep{Fu+17}. The metallicities are set to the solar values. The column density $N_{\rm H}$, the ionization parameter $\xi$, and the redshift $z$ of the absorbing gas are left as free parameters. Our assumption here is that each WA component is only sampling one distinct slab in the outflowing material along our line of sight.  Finally the XSTAR models of these WAs are generated, and the parameters used for these XSTAR models are listed in Table \ref{tab:tab3}.

We also attempt to derive constraints of the covering factor by the WAs. The compact accretion disk is generally considered fully covered by the WAs, whereas the BLR may not be completely covered by the WAs. Therefore, when fitting the COS spectrum, for accretion disk the covering factor is 1. For BLR it is determined by spectral fitting, and the XSPEC model {\tt partcov} is used to mimic this effect.

\begin{figure}
\centering
\includegraphics[angle=0,width=6in]{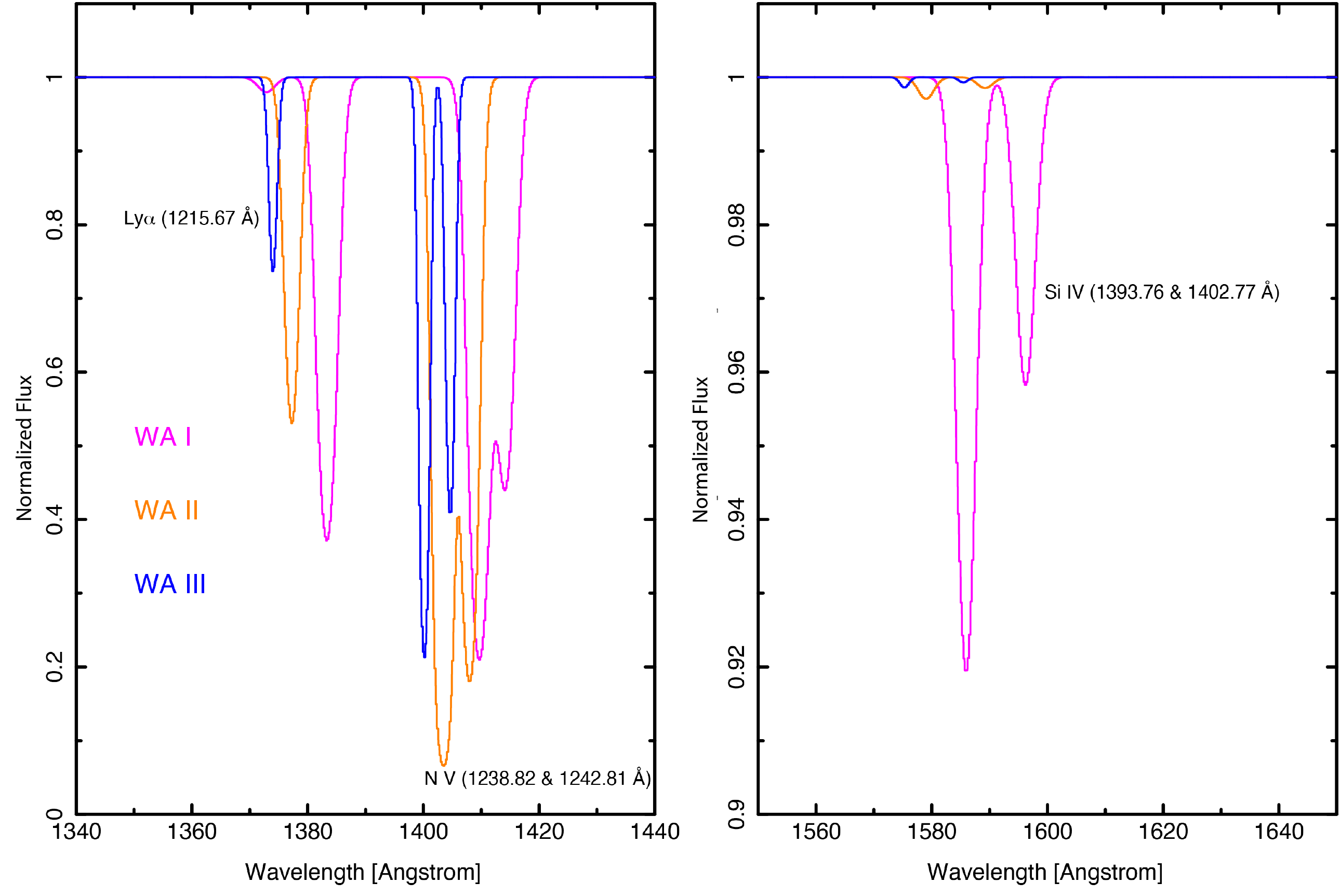}
\caption{The individual model components of the three WAs in PG 1001+054. The absorbers here are shown without the effect of the covering factor.
\label{fig:fig6}}
\end{figure}

\begin{figure}
\centering
\includegraphics[angle=0,width=7in]{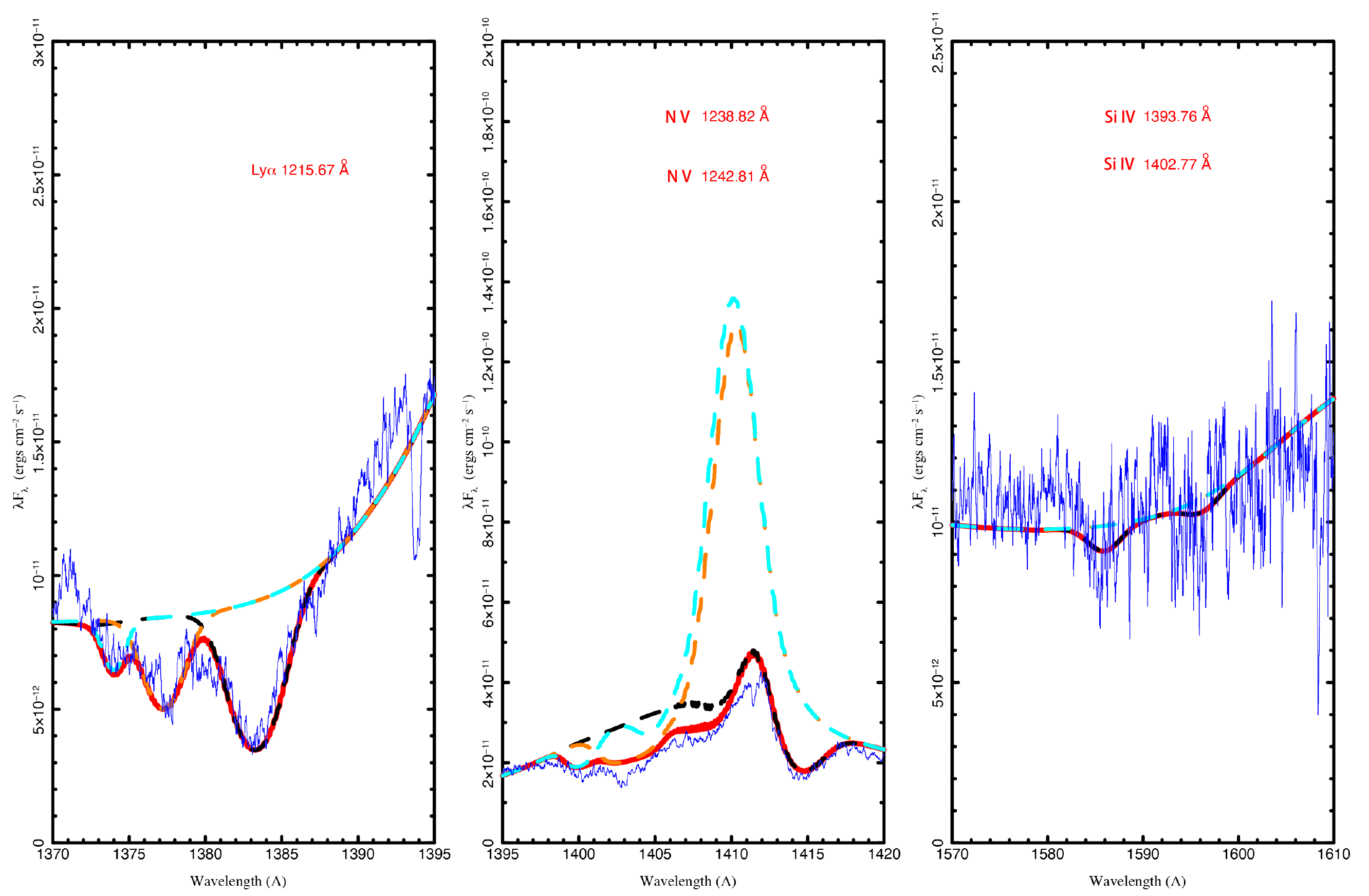}
\caption{The contribution of each absorber to the total multi-component absorption features in Ly$\alpha$, \nv~and \siiv~doublet. The contribution of WA I, II and III are represented by black, orange and cyan dashed line, respectively.  The combined model absorption is represented by the red solid line. The absorption of \nv~is mainly contributed by WA I, whereas WA II and III mainly produce absorption at the bluer part. Note that in the middle panel, \nv~is superposed on Ly$\alpha$ emission line, which is not significantly suppressed by absorption from these two WAs.
\label{fig:fig8}}
\end{figure}

\begin{figure}
\centering
\includegraphics[angle=0,width=4.5in]{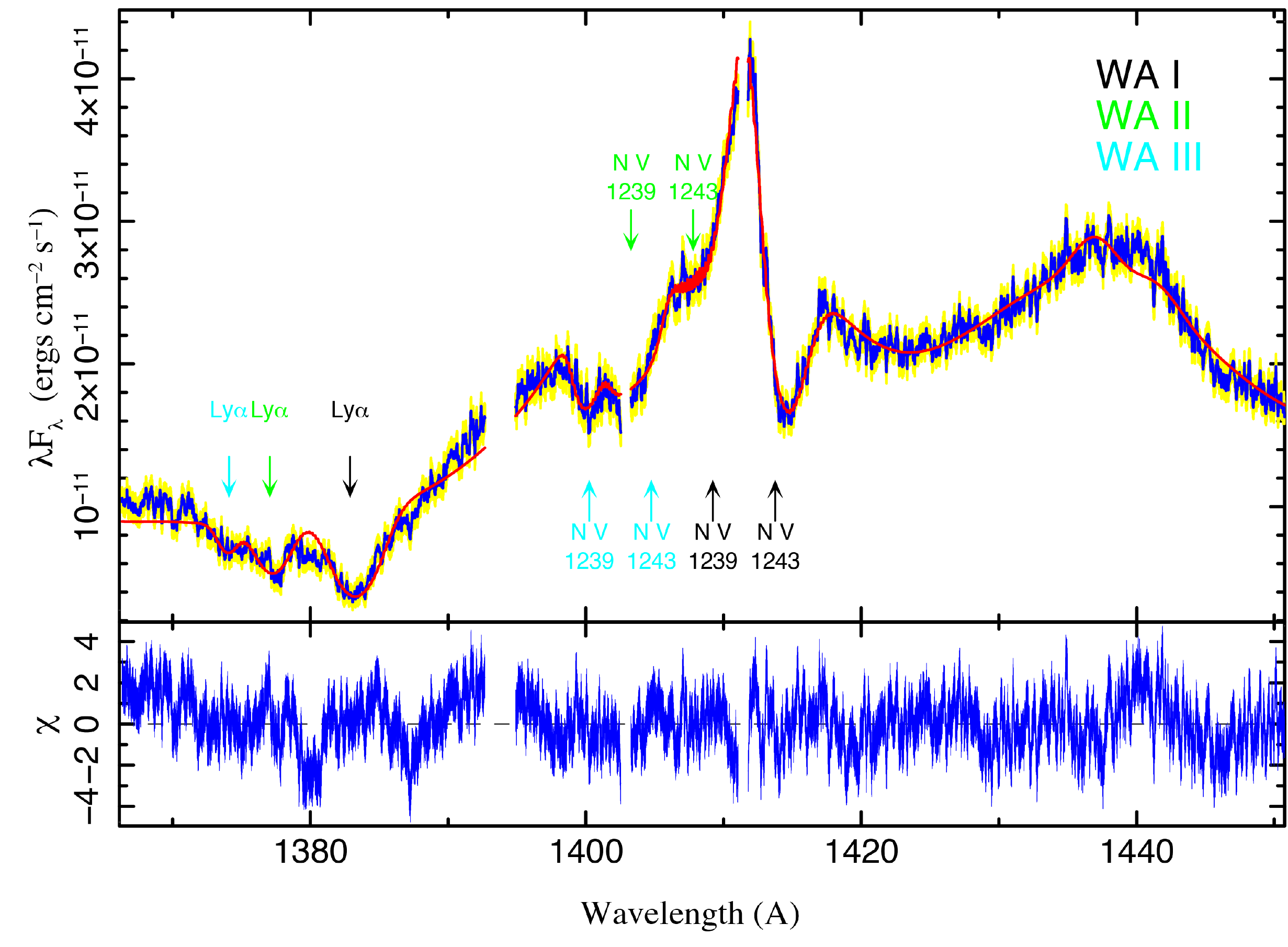}
\includegraphics[angle=0,width=4.5in]{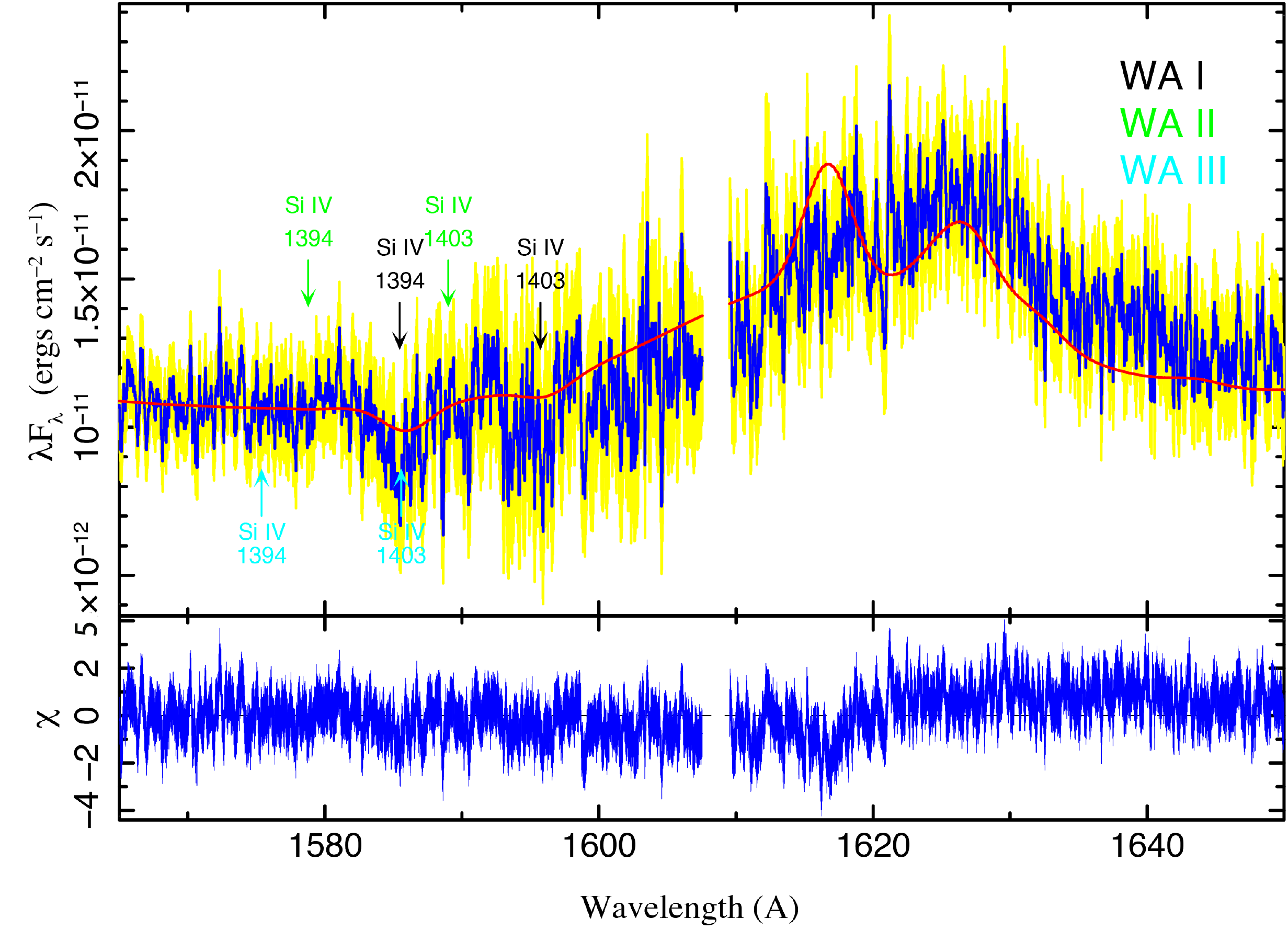}
\caption{The best-fit XSTAR models for Ly$\alpha$, \nv~and \siiv~absorptions in PG 1001+054, plotted over the observed {\em HST}/COS spectrum. The positions of WAs are labeled with arrows, and the ISM lines are masked out to prevent interference. Note that although the XSTAR model fit still leaves residuals at some emission lines in this complicated spectrum, we choose not to invoke more component. 
\label{fig:fig7}}
\end{figure}

Figure \ref{fig:fig1} shows the observed COS spectrum and the best-fit model including continuum, line emission, and absorption. We manually added a Gaussian absorption model component to account for the local broad Ly$\alpha$ absorption and a Gaussian emission component for the local \oi~ doublet (1302 $\textrm{\AA}$ and 1306 $\textrm{\AA}$) emission. The observed spectrum with complex line profiles appear well fitted by these physical components.  The photon index of the underlying power-law is $\Gamma$ = 3.20. A total of three components of WAs are used to describe the identified absorption lines in the observed COS spectrum. Table \ref{tab:tab4} lists the parameters of the best-fit WAs in PG 1001+054. Synthetic spectral models illustrating these components are shown in Figure \ref{fig:fig6}. All of the three WAs have noticeably high blue-shifted velocities, ranging from $\sim$6600 to $\sim$8900 \kmps. WA I has the lowest velocity, column density and ionization parameter among the three WAs, suggesting that it may be the furthest away from its central black hole. WA II and III have similar higher ionization states, larger column densities, and their velocities are higher than WA I. In Figure \ref{fig:fig8} we show the detailed contribution of the absorption lines in each WA (Ly$\alpha$, \nv~and \siiv~doublet) to the total multi-component absorption features. As shown in Figure \ref{fig:fig7}, the best-fit XSTAR models for Ly$\alpha$, \nv~and \siiv~absorption in PG 1001+054 are compared to the observed COS spectrum.


\begin{table}
\begin{center}
\caption{Parameters of the Emission Line Clouds and Identified Warm Absorbers.
\label{tab:tab4}}
\footnotesize
\begin{tabular}{ccccc}
\hline\noalign{\smallskip}
   & log$\xi$       & $N_{\rm{H}}$               & Redshift     &Covering \\
               &$\rm{(erg~s^{-1}~cm)}$  &$\rm{(\times10^{19}~cm^{-2})}$       &    & factor \\
\hline\noalign{\smallskip}
\hline\noalign{\smallskip}
Emission\\
\hline\noalign{\smallskip}
BLR I &$3.00^{+0.09}_{-0.01}$ & - &$0.160765^{+0.000007}_{-0.000051}$ &-\\
BLR II &$0.90\pm0.01$ &- &$0.159983^{+0.000014}_{-0.000001}$ &-\\
\hline\noalign{\smallskip}
Intrinsic Absorption\\
\hline\noalign{\smallskip}
WA I       &$0.65\pm0.02$  &$6.48^{+0.14}_{-0.18}$    &$0.137902^{+0.000007}_{-0.000018}$   &0.91 \\
WA II       &$1.90\pm0.02$  &$79.31^{+4.01}_{-4.17}$      &$0.132926^{+0.000025}_{-0.000029}$   & 0.12\\
WA III	      &$2.05\pm0.02$   &$32.98^{+1.57}_{-3.42}$  &$0.130239^{+0.000026}_{-0.000029}$   & 0.02 \\
\noalign{\smallskip}\hline
\end{tabular}
\end{center}
\end{table}

\section{Discussion} \label{sec:discuss}

\subsection{Comparison with Previous Work}

Being an X-ray weak BAL quasar \citep{Brandt+etal+2000,2014ApJ...794...70L,2022ApJ...936...95W}, there are few previous studies on the absorbers of PG 1001+054. \citet{Schartel+etal+2005} found that the column density $N_{\mathrm{H}}$ and ionization parameter $\xi$ of outflows in this source is $19.2\times10^{22}$~\cmsq and 542 $\rm~{erg~s^{-1}~cm}$ from its {\em XMM}-Newton observations. However, the column density $N_{\mathrm{H}}$ and ionization parameters $\xi$ of the absorber that we measured in the UV band are much lower than the X-ray derived values. It is plausible that the absorbers detected in the two bands are at different locations. 

For absorption system, it is notoriously difficult to obtain the location of absorbing gas along the line-of-sight. We attempt to estimate the distance of WAs away from the central black hole according to $\xi =L_{ion}/(nr^{2})$. $L_{ion}$ is the luminosity in the 1-1000 Ryd energy range, $n$ is the hydrogen number density, and $r$ is the distance to the ionizing source.  Without absorption lines from excited states to constrain on density, we cannot determine the exact value of $n_{\mathrm{H}}$ for these WAs in this source, nevertheless the $n_{\mathrm{H}}$ values for similar absorbers in other AGNs are generally larger than $10^{3}~\rm{cm^{-3}}$ \citep{Gabel+2005, 2018ApJ...865...90M,Aalto+2020, Arav+2013, Arav+2015, Arav+2020}. Due to the higher velocity of the outflowing absorbers in this source than those of typical WAs, it is likely that the $n_{\mathrm{H}}$ may also be slightly larger, hence a lower limit of $n_{\mathrm{H}}$ as $10^{3}~\rm{cm^{-3}}$ is adopted. On the other hand, previous work studying BAL/mini-BAL outflows frequently take $r$ = 1 pc as a placeholder radial distance to derive relevant properties (\citealt{Hamann+2019} and references therein). Following this approach we find the corresponding $n_{\mathrm{H}}$ is between $2\times10^{5}-5\times10^{6}~\rm{cm^{-3}}$ for our outflows. Using these values, we obtained that a plausible range for the location of these absorbers is between 1 and 73 pc. Table \ref{tab:tab5} lists the detailed information of the radial distance of these WAs from the nucleus. The location of these WAs implies that these WAs may be possibly linked to the ISM beyond the torus scale.


\begin{table}
\begin{center}
\caption{The Derived Properties of the Identified WAs in PG 1001+054.
\label{tab:tab5}}
\footnotesize
\begin{tabular}{ccccccccc}
\hline\noalign{\smallskip}
& log$\xi$  & $N_{\rm{H}}$  &  $v_{\rm out}$  &$n_{\rm e}$ & $r$ & $\dot{M}_{\rm out}$ & $\dot{E_k}$ &$\dot{E_k}/L_{Edd}$ \\
  &$\rm{(erg~s^{-1}~cm)}$  &$\rm{(\times10^{19}~cm^{-2})}$  &$(\kmps)$    &$(\mathrm{cm^{-3}})$&(pc) &$\rm{(M_{\odot}\, yr^{-1})}$ &$\rm{(erg~s^{-1})}$ &\\
\hline\noalign{\smallskip}
\hline\noalign{\smallskip}
WA I  &$0.65$  &$6.48$  &$-6660.73$ &$10^{3}-5\times10^{6}$ &1-73 &0.05-4.00 &$7.36\times10^{41}-5.57\times10^{43}$ & $0.013\%-0.967\%$\\
WA II  &$1.90$  &$79.31$ &$-8152.49$ &$10^{3}-3\times10^{5}$&1-17 &0.11-1.87 &$2.26\times10^{42}-3.92\times10^{43}$  &$0.039\%-0.681\%$\\
WA III	&$2.05$ &$32.98$  &$-8958.02$  &$10^{3}-2\times10^{5}$ &1-15 &0.01-0.12 &$2.08\times10^{41}-3.03\times10^{42}$  &$0.004\%-0.053\%$\\
\noalign{\smallskip}\hline
\end{tabular}
\end{center}
\end{table}

It is also worth noting that the location of absorbers observed in the UV band appears to be slightly further from the central black hole than that observed in the X-ray band. \citeauthor{Wang+etal+2000} (2000) also reached the same suggestion, finding that the X-ray absorbers require a column densities of at least a few times $10^{22}$~\cmsq ~in this source, which are much larger than that inferred from the UV absorption lines. They compared the {\em ROSAT} X-ray data and the UV absorption and emission lines and suggested that the observed UV line optical depth is much lower than expected from the X-ray absorbing column density. Our results are consistent with their conclusions. 

\begin{figure*}
\gridline{\fig{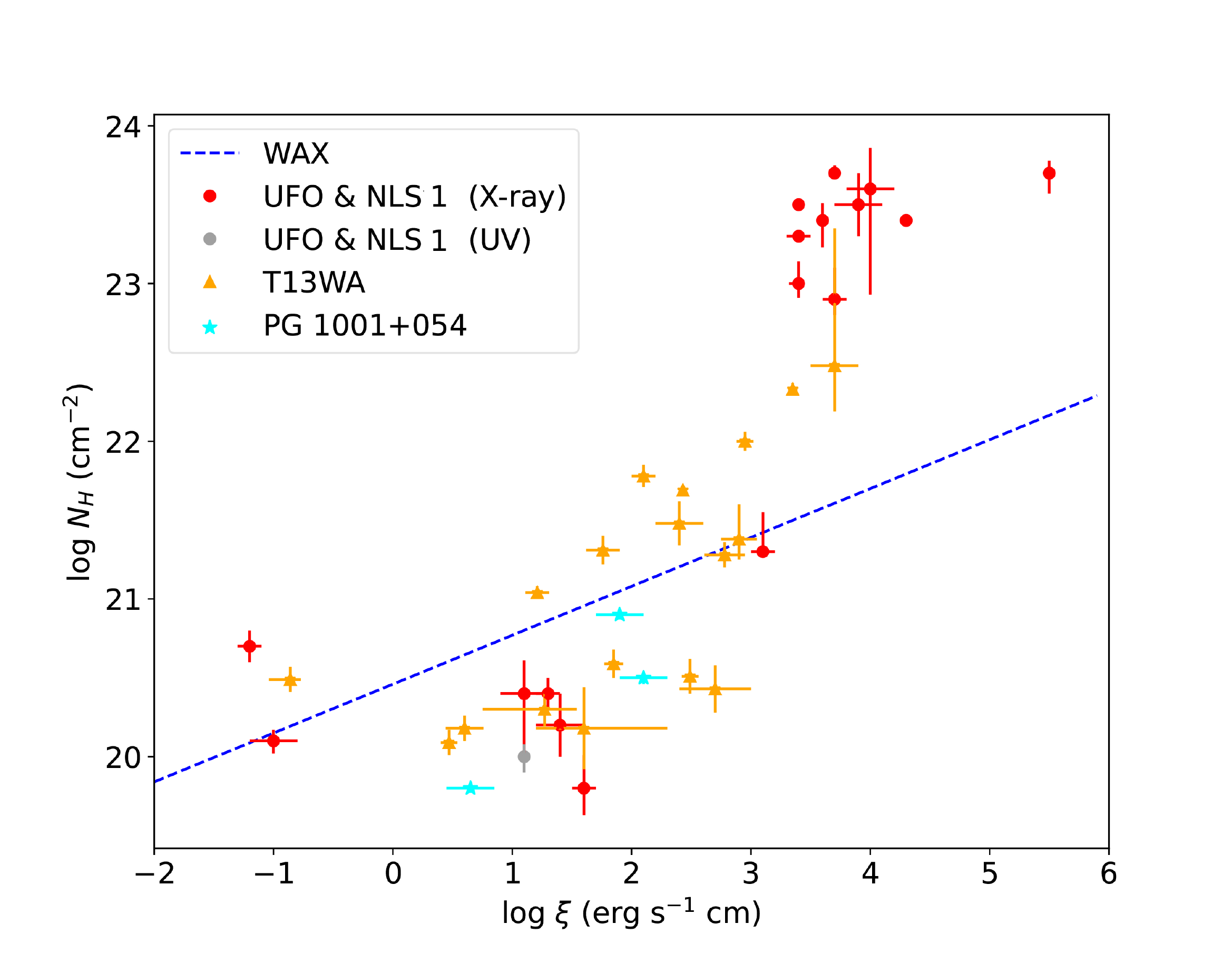}{0.51\textwidth}{(a)}
          \fig{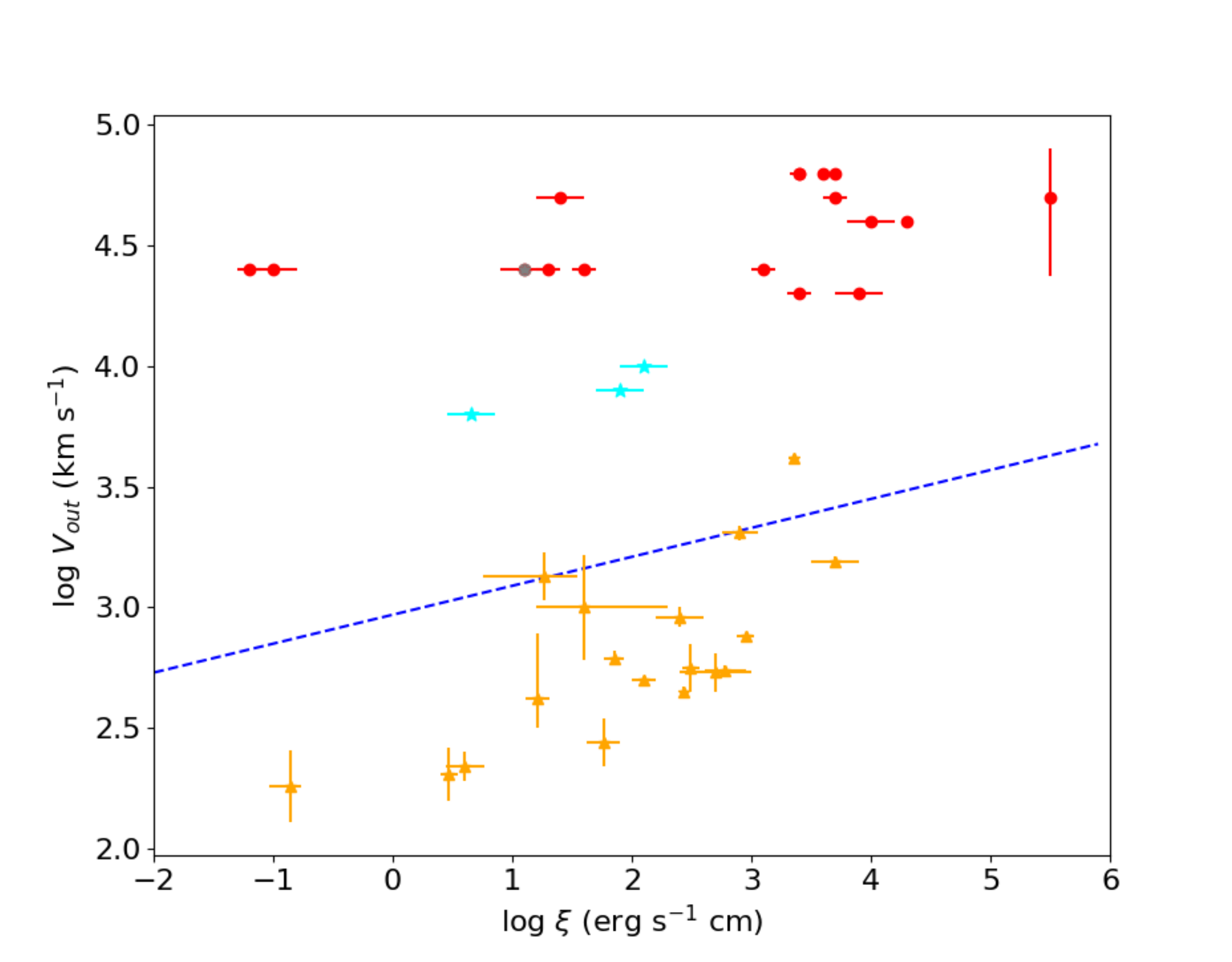}{0.51\textwidth}{(b)}}
\gridline{\fig{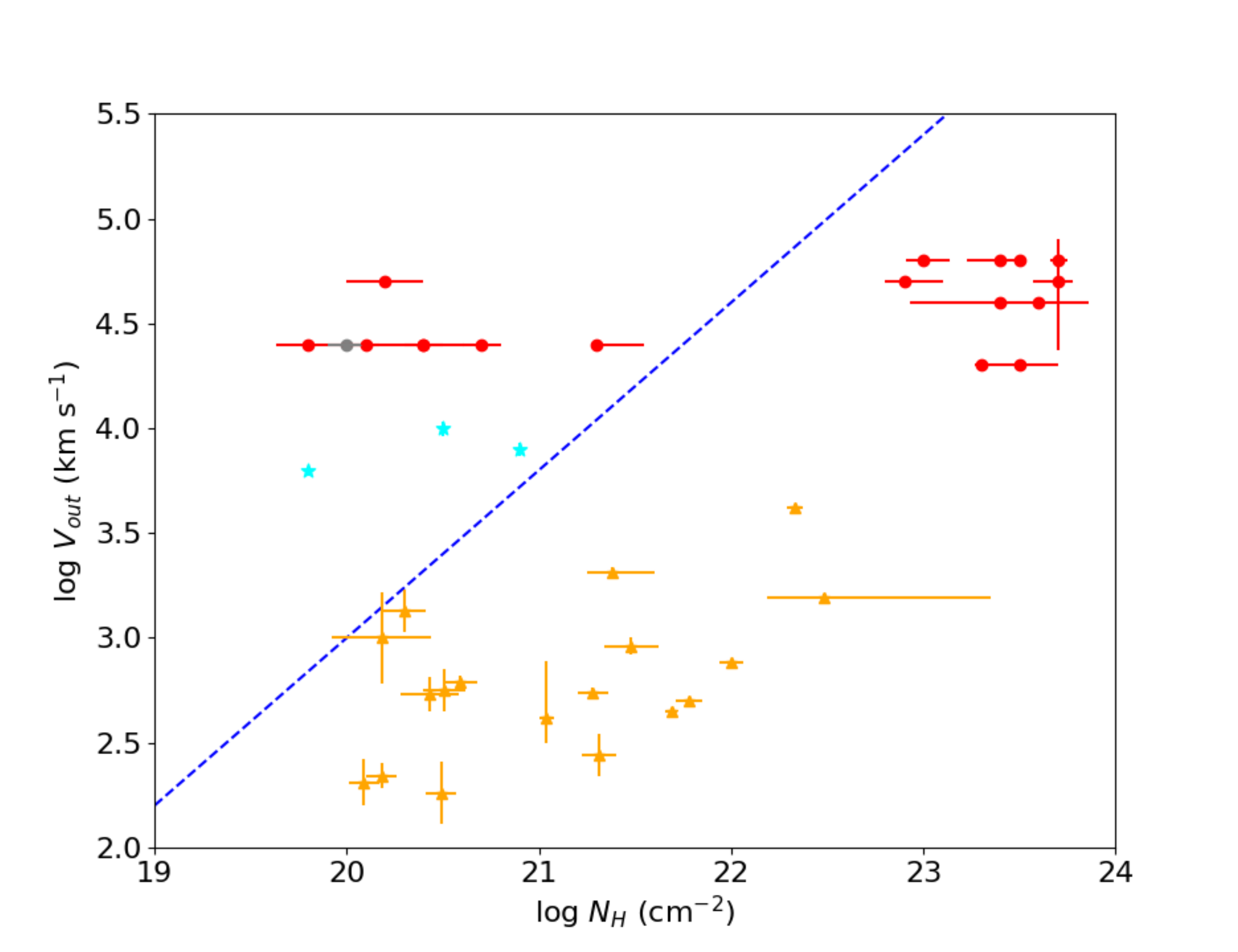}{0.51\textwidth}{(c)}}
\caption{Comparison of parameters between PG 1001+054 and different outflows compiled from literature, shown as scattered plots. (a) log$\xi$ versus log$N_{\rm{H}}$; (b) log$\xi$ versus log$v_{\rm out}$; (c) log$N_{\rm{H}}$ versus log$v_{\rm out}$. The symbols are derived values for PG 1001+054 (light blue star), linear fits for WAX sample from \citet{Laha+etal+2014} (blue dashed line), UFOs in the NLS1s in the X-ray (red circle) and the UV (grey circle) bands, and WA sample from \citet{Tombesi+etal+2013} (orange triangle). The error bars indicate the lower and upper limits of the parameters if reported in previous studies.
\label{fig:fig12}}
\end{figure*}

In a broader context, the outflowing UV absorber identified in PG 1001+054 can be compared to a well studied sample of UV/X-ray absorbers in local AGN and PG quasars. For example, \citet{Laha+etal+2014} carried out a homogeneous analysis of Warm absorbers in X-rays (WAX) in 26 Seyfert galaxies using {\em XMM}-Newton spectra and performed linear regression fits for the WA parameters that could be compared as typical values. The seminal work by \citet{Tombesi+etal+2013} also provided a sample of soft X-ray WAs of type 1 Seyfert galaxies. In addition, from the literature we compiled the derived properties for a list of NLS1 sources similar to PG 1001+054 that have reported UFO measurements, either in the X-ray or in the UV.  These include IRAS 17020+4544 (\citealt{Mehdipour+etal+2022} and \citealt{Sanfrutos+etal+2018}), Mrk 1044 (\citealt{Krongold+etal+2021}), PG 1211+143 (\citealt{Pounds+etal+2016}), 1H1934-063 (\citealt{Xu+etal+2022}), PG 1448+273 (\citealt{Laurenti+etal+2021}), 1H 0707-495 (\citealt{Kosec+etal+2018}), IRAS 13224-3809 (\citealt{Jiang+etal+2022}), and Mrk 590 (\citealt{Gupta+etal+2015}). In Figure \ref{fig:fig12} we show the distribution of different outflow parameters for these sources together with our measurements for PG 1001+054. Overall the data points are rather scattered in the parameter space, with the exception of the ionization parameter $\log \xi$ versus column density $\log N_{\rm H}$. Note that no fit for correlation analysis is intended here. \citet{Laha+etal+2014} previously identified there is such a correlation.  Figure~\ref{fig:fig12}a shows that the UV absorber in PG 1001+054 is typical of the WAs in ionization and column density, but in terms of outflow velocity, it is between UFOs ($v\sim 0.1c$) and other typical WAs ($v\sim 1000$ km s$^{-1}$). The significantly higher outflow velocity ($v \sim 7000-9000$ km s$^{-1}$) implies that a fast outflow is clearly present in this NLS1. This is consistent with the expectation that launching mechanism of UFOs is related to the radiatively driven wind \citep{Matzeu+17,M+22} and that NLS1s have overall high Eddington ratios.

Together with the detection of high ionization absorber in the X-ray for PG 1001+054 \citep{Schartel+etal+2005}, we suggest that the fast outflowing UV absorber seen in COS spectrum is probably part of a multiphase outflow. In the X-ray obscuration scenario, \citet{2022ApJ...936...95W} suggest powerful high density wind launched from the accretion disk. Most likely the low resolution and low S/N X-ray spectra due to its nature of an X-ray weak quasar prohibited a firm detection of any X-ray UFOs in this source. Figure \ref{fig:fig12} illustrates that the parameters of outflows span a wide range in ionization states, column densities, and velocities, which are consistent with the characteristics of stratified winds. Several AGN were detected with multi-component UFOs in terms of both ionization and velocity (e.g., IRAS 17020+4544,\citealt{Mehdipour+etal+2022}; PDS 456, \citealt{Reeves+2018}), showing Ly$\alpha$ UFO features besides the X-ray absorption signatures (e.g., Fe~XXV, Fe~XXVI, O~VII etc.). As discussed in \citet{Krongold+etal+2021}, such structure is likely produced by the outflow launched from AGN at the accretion disk scale shocks the ambient ISM \citep{2015ARA&A..53..115K}.  

\subsection{Mass and Energy Outflow Rate}

Bi-cone is generally adopted as a natural geometry for WA outflows \citep{Dorodnitsyn+etal+2008}. Following the formula for calculating mass loss rate \citep{Tombesi+etal+2015} $\dot{M}_{\mathrm{out}} =4\pi\mu r N_{\mathrm{H}}v_{\mathrm{out}} m_{\mathrm{p}}C_{f}$, where $m_{\mathrm{p}}$ is the mass of the proton, $\mu$ the mean atomic mass per particle ($\mu$= 1.4),  $v_{\mathrm{out}}$ the line-of-sight outflow velocity, $r$ the distance from the central black hole, and $C_f$ is the covering factor of each WA. According to the result in Table \ref{tab:tab5}, we estimate that the mass outflow rate ($\dot{M}_{\rm out}$) caused by these WAs is about 0.2-6 $M_{\odot}$ per year. Table \ref{tab:tab5} lists the detailed information of the mass outflow rate of each outflowing absorber. The energy outflow rate, or kinetic luminosity, is estimated following $\dot{E_k}=1/2 \dot{M}_{\mathrm{out}}v_{\mathrm{out}}^{2}$ (\citealt{Krongold+etal+2021}). The detailed energy outflow rates caused by each WA are listed in Table \ref{tab:tab5}.

We take the mass of SMBH in PG 1001+054 from \citet{Marian+etal+2020}, $M_{BH}=10^{7.7}M_{\odot}$, and the corresponding Eddington luminosity of PG 1001+054 is $L_{Edd}= 1.5\times 10^{12} L_{\odot}$. Summarizing contribution from each WAs, we obtain the ratio between total $\dot{E_k}$ and $L_{Edd}$, which ranges between 0.1$\%$ and 1.7$\%$. Theoretical models for efficient AGN feedback typically require the kinetic luminosity of AGN outflows is at least 0.5\%--5\% of the Eddington luminosity to have a significant impact on the galaxy evolution \citep[e.g.,][]{2005Natur.433..604D,Hopkins+etal+2010}. If located at tens of parsecs, the kinetic power carried by the WA outflows could be sufficient for a significant impact on the host galaxy of PG 1001+054 under this criterion. Future observations of the host properties may further reveal evidence for effective feedback.

\section{Summary} \label{sec:summary}

In this work, we analyze the {\em HST}/COS spectra of NLS1 galaxy PG 1001+054, and report on the discovery of fast outflowing UV absorbers. We fit the high resolution UV spectrum of PG 1001+054 and perform photoionization models, taking into account the physical components of local gas absorption, local dust extinction, Comptonized corona emission of the AGN, black body emission from the accretion disk, BLR emission, and absorption due to intrinsic WAs. The main findings are summarized as follows.

\begin{itemize}
\item The UV outflow is seen as narrow and blueshifted absorption lines of Ly$\alpha$, \nv, and \siiv, with velocities in the range of $\sim$7000-9000 km s$^{-1}$. The presence of three WAs can explain well the significant absorption lines in the COS spectrum and we derive their physical properties. A fast outflow is clearly present in this NLS1, and consistent with the expectation that launching mechanism of outflows is related to radiatively driven wind and that NLS1s have overall high Eddington ratios.

\item The possible location of WAs can be estimated to range between $\sim$1 pc and $\sim$73 pc away from the central black hole, which implies that these WAs may originate at or beyond the torus scale. The UV absorber in PG 1001+054 is typical of the WAs in ionization and column density, but shows significantly higher outflow velocity than other WAs ($v\sim 1000$ km s$^{-1}$). Together with previous detection of high ionization absorber in the X-ray for PG 1001+054 \citep{Schartel+etal+2005}, we suggest that the fast outflowing UV absorber seen in COS spectrum is probably part of a multiphase outflow. The parameters of outflows are consistent with the characteristics of stratified winds. Such structure is likely produced by the outflow launched from AGN at accretion disk scale shocks the ambient ISM.  

\item We estimate a total mass outflow rate of the three WAs $\dot{M}_{out}\approx 0.2-6~M_{\odot}$ yr$^{-1}$.  The ratio of total $\dot{E_k}$ of the WA outflows over $L_{Edd}$ is $\approx$0.1$\%$-1.7$\%$, which could potentially make sufficient influence on the evolution of the host galaxy when compared to the values in some theoretical models for efficient AGN feedback.

\end{itemize}

\begin{acknowledgments}

We sincerely thank the anonymous referee for the critical reading and helpful suggestions. We thank Dr. Shuinai Zhang and Mouyuan Sun for beneficial discussion, and Li Xue for technical assistance in computing. We also thank Jianfeng Wu, Bin Luo and Defu Bu for giving some useful advices. J.W. acknowledges support by the NSFC grants (U1831205, 12033004, 12221003), and the science research grants from CMS-CSST-2021-A06 and CMS-CSST-2021-B02. This research is based on observations made with the NASA/ESA Hubble Space Telescope obtained from the Space Telescope Science Institute, which is operated by the Association of Universities for Research in Astronomy, Inc., under NASA contract NAS 5–26555. These observations are associated with HST program 13347.

\end{acknowledgments}

\vspace{5mm}
\facilities{HST (COS)}


\bibliography{aaspg1001}{}
\bibliographystyle{aasjournal}


\end{CJK*}
\end{document}